\documentclass[twocolumn,accepted=2018-05-25]{quantumarticle}
\usepackage{amssymb}
\usepackage{amsmath}
\usepackage[numbers]{natbib}
\usepackage{graphicx}
\usepackage{float}
\usepackage[pdfpagelabels,pdftex,bookmarks,breaklinks]{hyperref}
\usepackage{tikz}
\usepackage[all]{hypcap}
\hypersetup{colorlinks,citecolor=blue,urlcolor=blue,linkcolor=blue}

\begin{document}
\title{Halving the cost of quantum addition}
\author{Craig Gidney}
\affiliation{Google, Santa Barbara, CA 93117, USA}
\email{craiggidney@google.com}

\begin{abstract}
We improve the number of T gates needed to perform an $n$-bit adder from $8n + O(1)$ \cite{Amy2013, Cuccaro2004, AustinDiscussionsAndEmails2017} to $4n + O(1)$.
We do so via a ``temporary logical-AND" construction which uses four T gates to store the logical-AND of two qubits into an ancilla and zero T gates to later erase the ancilla.
This construction is equivalent to one by Jones \cite{Jones2013}, except that our framing makes it clear that the technique is far more widely applicable than previously realized.
Temporary logical-ANDs can be applied to integer arithmetic, modular arithmetic, rotation synthesis, the quantum Fourier transform, Shor's algorithm, Grover oracles, and many other circuits.
Because T gates dominate the cost of quantum computation based on the surface code, and temporary logical-ANDs are widely applicable, this represents a significant reduction in projected costs of quantum computation.
In addition to our $n$-bit adder, we present an $n$-bit controlled adder circuit with T-count of $8n + O(1)$, an out-of-place adder that can be uncomputed without using T gates, and discuss some other constructions whose T-count is improved by the temporary logical-AND.
\end{abstract}

\maketitle

\section*{Introduction}

The surface code \cite{Brav98,Denn02,Raus07,Raus07d,Fowler2012} is a quantum error correcting code that works on a 2D nearest-neighbour array of qubits and achieves a threshold error rate of approximately 1\%.
This makes the surface code a likely component in the architecture of future error corrected quantum computers, because 2D arrays of qubits with nearest-neighbor connections are possible with many qubit technologies \cite{Schl11,Bare13,Gamb17,Leik17,Laht17} and other well understood error correcting codes either have lower thresholds or require stronger connectivity.

\begin{figure}
  \includegraphics[width=\linewidth]{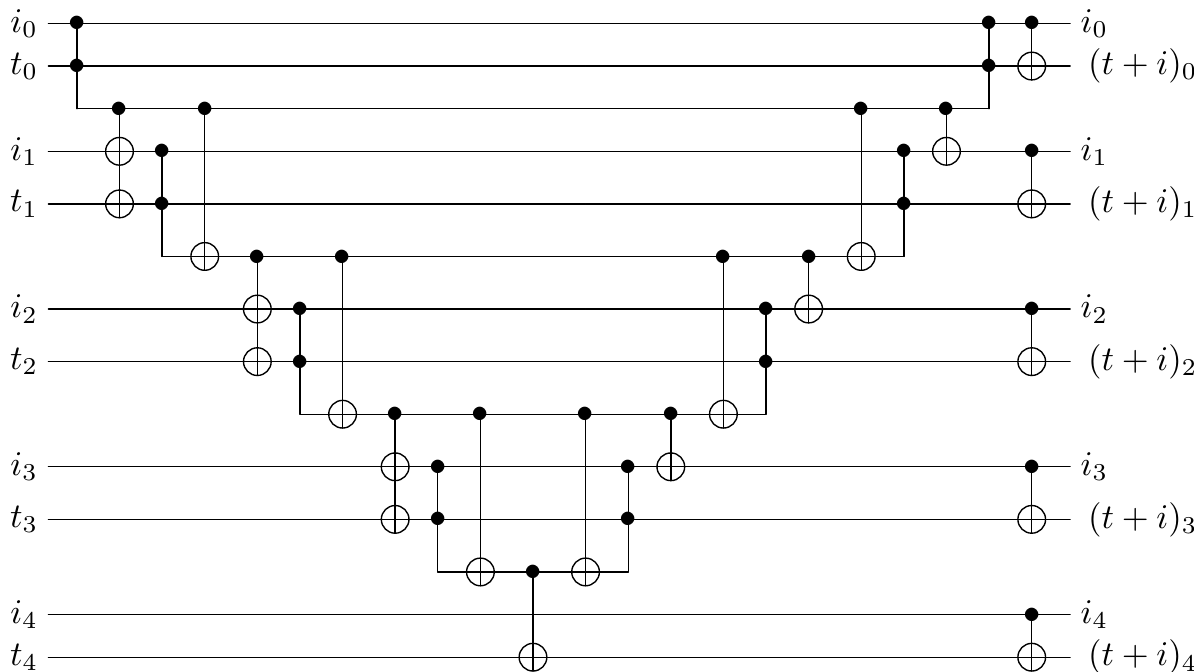}
  \caption{
	A 5-bit adder with T-count of 16.
	Uses Clifford operations, four logical-AND computations each with a T-count of 4, and four logical-AND erasures requiring no T gates.
	Generalizes to an $n$-bit adder with a T-count of $4n - 4$ and a measurement depth of $2n - 2$.
	See \autoref{fig:adder-building-block} for the adder building-block and \autoref{fig:temporary-logical-AND} for the logical-AND computation and uncomputation circuits.
  }
  \label{fig:adder}
\end{figure}

One of the downsides of the surface code is that it has no cheap mechanism to apply non-Clifford operations such as $T$ gates.
Instead, $T$ gates are performed by distilling and consuming $|T\rangle = \frac{1}{\sqrt{2}} (|0\rangle + e^{i \pi/4} |1\rangle)$ states.
The T gate is performed via a controlled-NOT from the target qubit into the $|T\rangle$ ancilla, followed by a classical feedback step that measures the ancilla to determine if an $S$ gate is applied to the target.
Unfortunately, distilling the $|T\rangle$ state that will ultimately be passed into this process has significant cost.
The cost of distillation is high enough that error corrected quantum computations are expected to bottleneck waiting for $|T\rangle$ states to be produced, so that their runtime is dominated by the number of T gates (as opposed to being dominated by gate count, circuit depth, or even measurement depth).
For example, although our adder has the same measurement depth as the Cuccaro adder (i.e. $2n + O(1)$), we still expect our adder to execute more quickly in practice on early error corrected machines (assuming there is room for the $n$ ancillae our adder requires).

Because the surface code is a likely component of future quantum computers, and the runtime of computations within the surface code will be dominated in practice by the number of T gates, it is important to optimize the number of T gates used by quantum circuits.
Optimizing the T-count of basic elements of quantum circuits, such as the construction of adders and Toffoli gates, is particularly important because any improvement is widely applicable.

The textbook construction of a Toffoli gate uses seven T gates \cite{Nielsen2009}.
When Toffoli operations are paired, i.e. when an initial Toffoli operation is later uncomputed by a second Toffoli operation, each Toffoli in the pair can omit three of the T gates from the textbook construction.
This introduces phase errors but, assuming intermediate operations aren't sensitive to the phase errors, the second Toffoli gate can uncompute the phase errors while uncomputing the state permutation \cite{Barenco1995, Nielsen2009}.
It is also possible to reduce the T-count of an unpaired Toffoli gate to 4 by using an ancilla qubit and a classically conditioned fixup operation \cite{Jones2013}.

The Cuccaro adder uses $2n + O(1)$ Toffoli gates \cite{Cuccaro2004}.
Existing constructions implement the Cuccaro adder's Toffolis using $8n + O(1)$ T gates \cite{Barenco1995, Cuccaro2004, Amy2013}.
This T-count can be achieved either by applying the ancilla-and-fixup construction from \cite{Jones2013} to each individual Toffoli gate, or by noting that all but one of the adder's Toffoli gates appear in compute/uncompute pairs and then applying the matched-phase-error construction from \cite{Barenco1995, Nielsen2009} to each pair.

\begin{figure}
  \includegraphics[width=\linewidth]{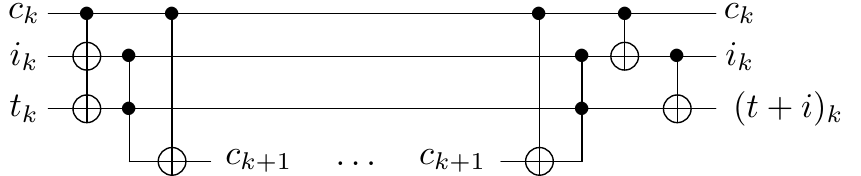}
  \caption{
	Our adder circuit building-block, with a T-count of 4 and a measurement-depth of 2.
	A variant of the Cuccaro adder building-block \cite{Cuccaro2004}.
  }
  \label{fig:adder-building-block}
\end{figure}

The leading factor of 8 in the T-count of an $n$-bit adder has stood for over a decade \cite{Barenco1995, Cuccaro2004, AustinDiscussionsAndEmails2017}.
We improve the leading factor with a construction based on the temporary logical-AND, combining ideas from both the matched-phase-error construction and the ancilla-and-fix-up construction.
We thereby halve the number of T gates needed to perform Toffoli gates that appear in compute/uncompute pairs and thus halve the T-count of Cuccaro-style adders from $8n + O(1)$ to $4n + O(1)$.

Although the title of this paper focuses on addition, we consider our main {\em conceptual} contribution to be reframing the ancilla-and-fix-up construction from \cite{Jones2013} into the temporary logical-AND operation.
This conceptualization makes it clear that the construction is more widely applicable than previously realized.
Anytime operations share controls, even if those operations are far apart, a temporary logical-AND allows the controls to be combined once instead of once per operation (as long as intermediate operations are not sensitive to phase error in the controls).
Addition is just one example of an operation where this is beneficial.
We also discuss other examples.

\section*{Opportunity Cost of Ancillae}

Our adder construction consumes $4n$ fewer $|T\rangle$ states than the Cuccaro adder, but in doing so it holds $n$ ancillae through a measurement-depth of $2n$.
Before we discuss our adder construction, we must first discuss the question: is that tradeoff worth making?

Obviously, if performing an addition on a small quantum computer with no space to spare for additional ancillae, it would be not be possible to use our adder whereas Cuccaro's would still be viable.
However, lack of space is not the only reason ancillae can be problematic.
In particular, note that any qubits being used as ancillae are qubits {\em not being used in T factories}.

It is possible to produce a high-quality $|T\rangle$ state in 960 units of ``spacetime volume" \cite{Babbush2018} (though this is certainly not a lower bound!).
For comparison, an ancilla qubit stored as a double-defect through $k$ serial measurements will cover at least $2k$ units of spacetime volume.
Putting these two facts together, and ignoring the fact that T factories are discrete objects with packing constraints, we will approximate the opportunity cost of holding an ancilla as $\frac{1}{480} |T\rangle$ states per measurement-depth.

With the above in mind, the ``effective" T-count of our adder, including the opportunity cost of ancillae (and accounting for the fact that some are kept for longer than others), is $\frac{1}{480} n^2 + 4n$ whereas the effective T-count of the Cuccaro adder is $8n$.
This implies that our adder is worse than the Cucarro adder when $n > 1920$.
However, switching adders at $n=1920$ ignores the possibility of using a hybrid of both adders where bit positions below a cutoff propagate carries via ancillae as in our adder and bit positions above the cutoff propagate carries inline as in the Cuccaro adder.
In a hybrid adder of this type, the cutoff actually occurs at $n=960$.

We caution that the cutoff of 960 that we just estimated is not a fixed constant.
It is affected by other optimizations, by future research, and by overall system design.
As the spacetime volume of T factories is improved, the cutoff {\em will} move downward.
Conversely, if the spacetime volume of ancillae is improved, the cutoff moves upward.
For example, qubits that are idle can be compressed into an inactive ``memory" form that covers six times fewer physical qubits than the double-defect form \cite{horsman2012}.
This means that {\em idle} ancillae, like the ones in our adder, have a lower opportunity cost than active ancillae.
This optimization increases the cutoff to $n \approx 5760$, which is beyond the size of adders that would be used to break 4096-bit RSA keys with Shor's algorithm.

An interesting consequence of ancillae having a T-count opportunity cost, besides making circuits that execute inplace desirable, is that it makes circuit depth relevant to the T-count.
Logarithmic-depth adders may use a constant factor more $|T\rangle$ states and ancillae than our adder, but they use exponentially less {\em ancilla-depth}.
The T-count opportunity cost of logarithmic-depth adders grows like $\Theta(n \lg n)$ instead of like $\Theta(n^2)$.
So, despite their ``raw" T-count being larger, for sufficiently large $n$ their effective T-count must become lower than the effective T-count of our ripple-carry adder.

Beware that the previous paragraph only applies in the regime where there are enough physical qubits to support enough T factories to properly feed a logarithmic-depth adder.
In the early error corrected regime, when only a few T factories are available, the constant-factor penalty on the T-count of logarithmic-depth adders will result in a runtime that is longer than the runtime of ripple-carry adders (because they both have to wait for $|T\rangle$ states).
It is only as more and more T factories become available that the parallelism inherent to logarithmic-depth adders becomes accessible, overcomes the constant-factor T-count penalty, and triggers a gradual transition away from ripple-carry adders.

Whenever ancillae are used to save T gates, it is important to consider that there may be alternative uses of those ancillae that net even more $|T\rangle$ states (or other desirable resources).
These tradeoffs depend on overall system design.
Overall system optimization is important, but it is not the subject of this paper and so we will limit ourselves to introducing basic tools (e.g. the concept of a temporary logical-AND, as well as an adder that uses $4$ T gates and one ancillae per bit instead of $8$ T gates per bit) that future system architects can combine with other techniques on a case by case basis.

\section*{Results}

In \autoref{fig:adder}, we present a 5-bit adder with a T-count of 16.
It performs 4 temporary logical-ANDs, each with a T-count of 4.
All other operations are Clifford operations, with no T-count.

The building block of our adder is shown in \autoref{fig:adder-building-block}.
We construct $n$-bit adders by nesting $n$ copies of the building block inside of each other.
The outer-most and inner-most blocks (which act on the low bit and high bit respectively) are then specialized based on the fact that they either have no carry input or no carry output.

\begin{figure}
  \includegraphics[width=\linewidth]{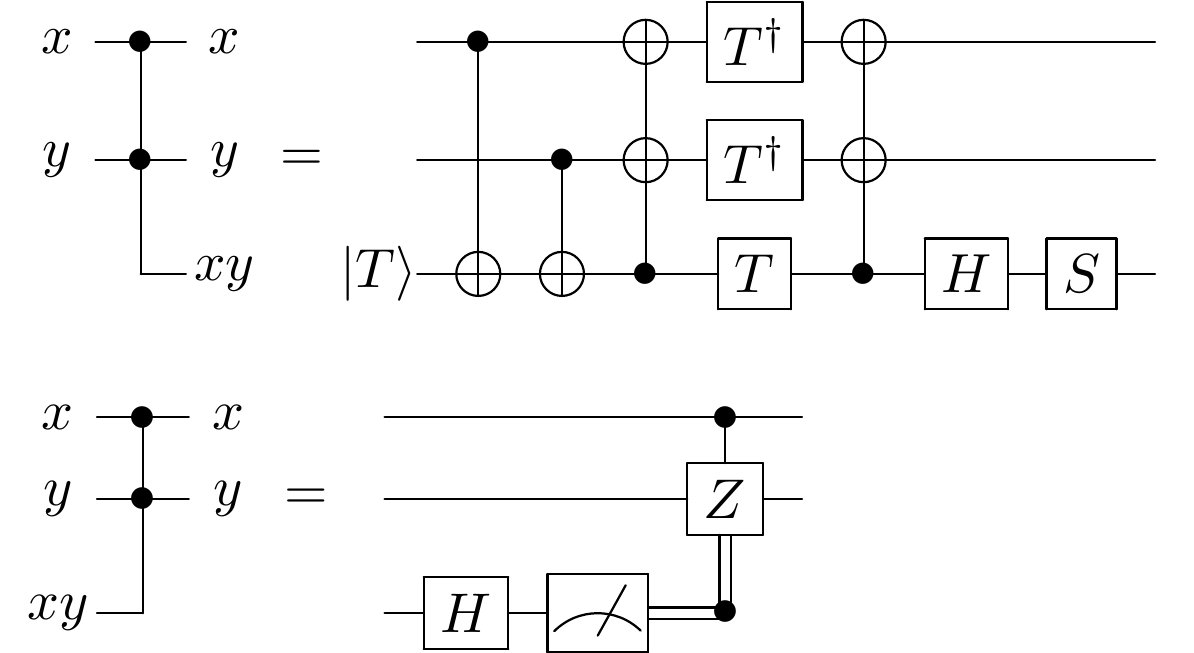}
  \caption{
	How to compute and uncompute the logical-AND of two qubits.
	The computation circuit (top) has a T-count of 4 and a measurement-depth of 1.
	For systems where $|T\rangle$ states cannot be used as an input without performing a measurement, it is still possible to achieve a measurement depth of 1 by rearranging the circuit and using a temporary ancilla (e.g. as in Figure 1 of \cite{Jones2013}).
	Note that the $|T\rangle$ state input contributes to the T-count, because $|T\rangle$ states are the resource used to perform T gates.
	The uncomputation circuit (bottom) uses a measure-and-fixup approach \cite{Jones2013} that requires only Clifford gates, and so has a T-count of zero and a measurement-depth of 1.
    \\
    An alternative uncomputation construction is to simply do the reverse of the computation circuit.
    This alternative approach has a net T-count of 2 (because a $|T\rangle$ state is recovered).
    The resulting temporary logical-AND with a T-count of 6 would still be an improvement on existing work, but would be inferior to the measure-and-fixup approach shown above.
  }
  \label{fig:temporary-logical-AND}
\end{figure}

Our adder uses temporary logical-AND operations, which we draw as wires emerging out of a pair of controls then later merging into the same pair of controls.
\autoref{fig:temporary-logical-AND} shows how we compute the logical-AND of the two controls, and also the corresponding uncomputation.

Computing the temporary logical-AND has a T-count of 4, but uncomputing it has a T-count of 0.
This surprising asymmetry is due to the fact that measurement is not reversible.
The uncomputation uses measurement in a way that the computation cannot.

\begin{figure}
  \includegraphics[width=\linewidth]{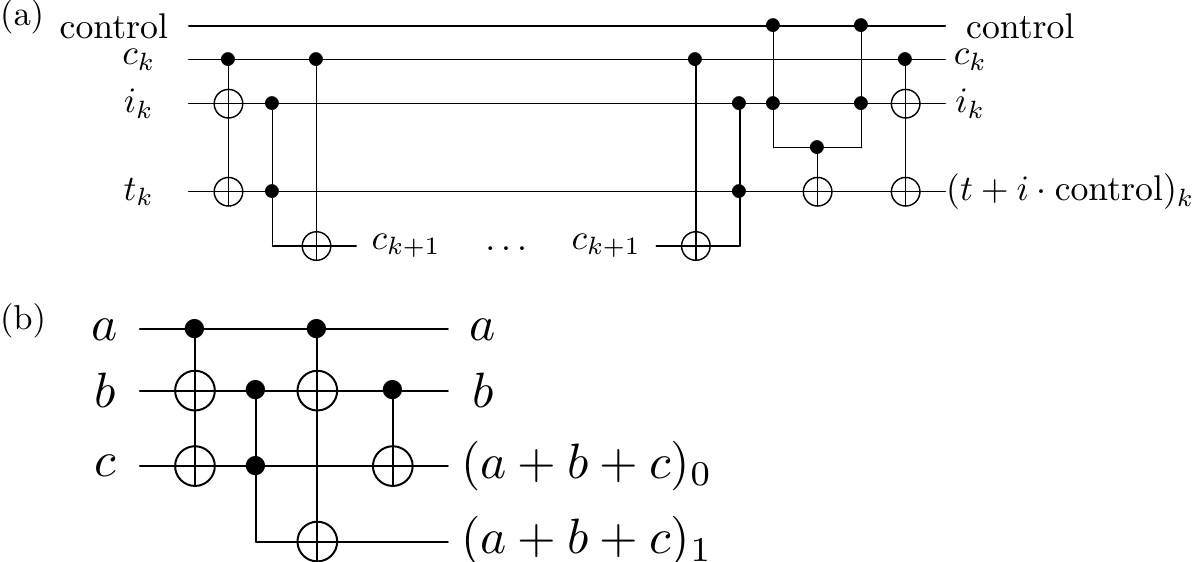}
  \caption{
	Variations on our adder construction.
	(a) is a controlled-adder building-block with a T-count of 8.
	The addition will only occur when the control is on.
	(b) is a standard out-of-place adder building-block, modified to use a temporary logical-AND.
	Because of this modification, the inverse of this building-block does not use any T gates.
	This means that uncomputing a register storing $a+b$, when registers storing both $a$ and $b$ are available, requires no T gates.
  }
  \label{fig:other-adder-building-blocks}
\end{figure}

\autoref{fig:other-adder-building-blocks} shows the building-blocks for two variations on our adder: a controlled adder and an out-of-place adder.

Some additions, such as the ones performed by the multiplications within the modular exponentiation in Shor's algorithm, are conditioned on a control qubit.
Our controlled adder reduces the cost of these additions from $21n + O(1)$ \cite{Coreas2017} to $8n + O(1)$.

Out-of-place adders are useful when a circuit is going to compute an addition, use it for awhile, then uncompute it.
Our out-of-place adder does not improve on the cost of computing an out-of-place addition.
However, because our out-of-place adder is based on computing a temporary logical-AND, its inverse does not use any T gates.
This makes it possible to uncompute an out-of-place addition while consuming no T gates.

Decreasing the T-count of addition reduces the T-count of any construction based on addition.
For example, in \cite{Fowler2012} it is estimated that factoring a 2048-bit number on a surface-code-based quantum computer would take $2 \cdot 10^{12}$ distilled $|T\rangle$ states and have a measurement-depth spanning 27 hours (though the actual computation would likely be bottlenecked on distillation rather than measurement).
The time estimate assumes Toffolis have a measurement-depth of 3, and the T-count estimate assumes Toffolis have a T-count of 7.
Shor's algorithm is dominated by the cost of controlled additions.
With our techniques, the average T-count and measurement-depth of the relevant Toffolis is $\sim 2.7$ and $1$ respectively.
With these numbers, the measurement-depth estimate is reduced to 9 hours and the T-count estimate is reduced to $8 \cdot 10^{11}$ distilled $|T\rangle$ states.

On top of reducing the T-count of obviously-related classical operations like multiplication and exponentiation, reducing the T-count of addition also reduces the T-count of quantum-specific operations such as rotating qubits.

For example, our improved adder allows the operation $R_Z(\theta)$ to be applied to $n$ qubits with a T-cost of $4n + O(\lg n \lg \frac{1}{\epsilon})$ as follows.
First, we must reduce the $n$ target qubits into a binary ``Hamming weight register", which indicates how many of the target qubits are \textsc{on}.
We start by assigning each qubit a ``weight" of 1.
We then start applying our out-of-place adder building-block to triplets of qubits of the same weight.
Each application of the adder will turn 3 relevant qubits of weight $w$ into 1 relevant qubit of weight $w$ and 1 relevant qubit of weight $2w$.
Note that this means the number of relevant qubits goes down by 1 per adder application, except when only 2 qubits of weight $w$ remain (which, as long as lower-weight qubits are always added first, occurs at most $\lg n$ times).
Since we start with $n$ qubits, and the Hamming weight register will have $\lg n$ qubits, and the two-remaining-qubits-of-same-weight case occurs at most $\lg n$ times, it will take at most $n - \lg n + \lg n = n$ adder applications to reduce the $n$ initial qubits into the $\lg n$ qubits forming the Hamming weight register.
Therefore, computing the Hamming weight register has a T-count of at most $4n$ (whereas uncomputing it will be free).
Now, for each position $p$ in the Hamming weight register, synthesize and apply the operation $R_Z(\theta \cdot 2^p)$ to the register qubit at that position.
This uses $O(\lg n \lg \frac{1}{\epsilon})$ T gates, which is negligible in comparison to $4n$ for large $n$.
Finally, uncompute the Hamming weight register (using no T gates).
This completes the application of the $n$ desired $R_Z(\theta)$ operations.

Another quantum operation that can be implemented via an adder is the $n$-qubit phase gradient operation $\text{Grad}_n = \sum_{k=0}^{2^n-1} e^{2 i \pi k / 2^n} |k\rangle \langle k|$.
Normally this operation would be implemented by separately applying the operation $R_Z(\pi 2^{-p})$ to each qubit of the target register, where $p$ is the qubit's index in the register and the number of T gates needed for each rotation depends on the maximum per-gate error $\epsilon$.
However, assuming a ``phase gradient register" prepared in the state $2^{-b/2} \sum_{k=0}^{2^b-1} e^{-2 i \pi k / 2^b} |k\rangle$ is available, the phase gradient operation can be performed via addition \cite{Kitaev2002}.
Add the target register into the phase gradient register, and phase kickback will apply the $\text{Grad}_n$ operation to the target.
With our adder, this construction performs phase gradients using $4n + O(1)$ T gates.
This T-count is interesting because it appears to be independent of $\epsilon$ despite the phase gradient operation involving arbitrarily small rotations.
However, there are three ways in which this phase gradient construction's cost does depend on $\epsilon$.
First, $\epsilon$ still bounds the minimum quality of the $|T\rangle$ states powering the T gates.
Second, large phase gradients can be truncated down to a size $n_{\text{max}}$ asymptotically equal to $\Theta(\lg \frac{1}{\epsilon})$.
Third, initializing the reusable phase gradient register has a one-time cost of $O(n_{\text{max}} \lg \frac{1}{\epsilon})$ T gates.

\begin{figure}
  \includegraphics[width=\linewidth]{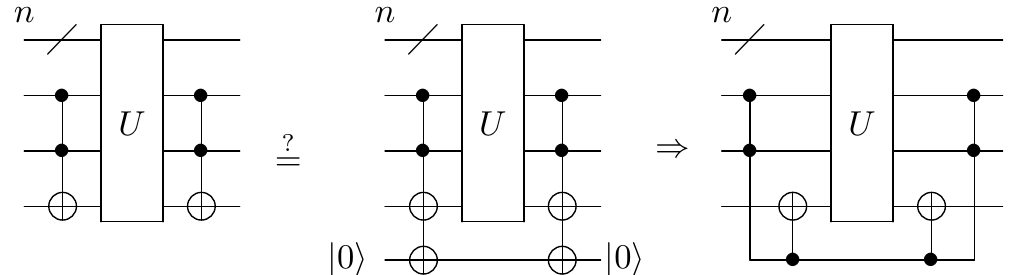}
  \caption{
	A sufficient condition for replacing a pair of Toffoli gates with a temporary logical-AND, saving 4 T gates.\\
	1) The later Toffoli gate must be uncomputing the earlier Toffoli gate.\\
	2) Intermediate operations must not be sensitive to the presence of the entangled ancilla.
  }
  \label{fig:paired-toffoli-replacement-rule}
\end{figure}

The temporary logical-AND is useful for optimizing an even wider variety of circuits than addition is.
Whenever Toffoli gates appear in compute/uncompute pairs, and intermediate operations are not sensitive to phase errors on the controls of the Toffoli gate (i.e. the condition in \autoref{fig:paired-toffoli-replacement-rule} is satisfied), it is possible to save 4 T gates by replacing the pair of Toffoli gates with a temporary logical-AND.

For example, a simple way to construct Grover oracles starts by translating a classical predicate into a classical reversible circuit made with Toffoli gates.
This reversible circuit is then used on a quantum computer to compute an output qubit, and a Z gate is applied to the output qubit before the circuit is run in reverse to uncompute the output qubit.
Every Toffoli gate generated by this construction is part of a compute/uncompute pair (each Toffoli in the computation will match a Toffoli in the uncomputation).
Using temporary logical-ANDs, instead of individually translating each Toffoli gate into T gates, halves the T-count of this approach to constructing Grover oracles.
However, note that this naive optimization tends to require an unreasonable number of ancillae and will not generalize to all methods for producing Grover oracles.

As another example, note that the temporary logical-AND can perform NOT gates with many controls by iteratively combining the controls down to a single representative ancilla.
The representative ancilla is then used to control a CNOT onto the target qubit.
This construction takes $4n-4$ T gates to perform a NOT with $n$ controls (and is equivalent to a nesting construction from \cite{Jones2013}).
As when replacing Toffoli gates in compute/uncompute pairs with a temporary logical-AND, the ancilla representing the $n$ controls should be kept and used multiple times whenever possible (instead of being uncomputed and recomputed).

Yet another example of a kind of circuit that benefits from the temporary logical-AND is {\em low-depth} adder circuits.
For example, the Toffoli gates produced by the $P$ and $P^{-1}$ rounds in Draper et al's logarithmic-depth adder \cite{Draper2004} form compute/uncompute pairs that can be replaced by temporary logical-ANDs.
Alternatively, by using temporary logical-ANDs to implement the classical Brent-Kung adder \cite{brent1982} in a reversible fashion, two $n$-bit numbers can be added in $O(\lg n)$ depth with a T-count of only $12n$.

Finally, we note that Toffolis also appear in compute/uncompute pairs in quantum circuits rooted in physics instead of mathematics.
For example, we used temporary logical-ANDs to cut the T-count of a chemistry algorithm in preparation by nearly 50\% \cite{Babbush2018}.

\section*{Discussion}

For over a decade, the T-count of addition has been $8n + O(1)$ \cite{Amy2013, Barenco1995, Cuccaro2004}.
In this paper we showed how to halve the leading factor of this cost by replacing Toffolis in compute/uncompute pairs with temporary logical-ANDs.
The temporary logical-AND is a basic circuit construction widget that can be used in many circuits, and a good example of measurement breaking the symmetry between computation and uncomputation (allowing one to be more efficient than the other).
We demonstrated how to optimize the T-count of a few tasks using our adder and the temporary logical-AND, but there are many other low hanging applications (e.g. converting between binary and unary, temporary sorting, applying an operation to a qubit indexed by a register in superposition, parameterized bit-rotation of a qubit register, computing the greatest common divisor of two values, etc).

It is interesting to consider if addition or temporary logical-ANDs can be done with still fewer T gates.
In \cite{Howard2017} it is proven that a Toffoli gate requires at least 4 T gates, though the specific nature of the proof does not eliminate the possibility that $n$ Toffolis could be implemented with fewer than $4n$ T gates for some large $n$.
Although we suspect the true lower bound really is 4, we also expect that there are regardless many opportunities to optimize how many T gates go into performing Toffolis.
For example, if a task involves repeating an action several times, T gates inside Toffolis at the end of one repetition could cancel out T gates in work being done at the start of the next repetition.

Regardless of whether there will be further T-count improvements for operations as basic as paired Toffoli gates, it is clear that many other kinds of T-count optimizations are waiting to be found.
Not just in the construction of basic low-level operations, but in medium-level constructions, high-level constructions, lower-than-circuit level constructions, and generally across the whole technology stack that will be needed to perform error corrected quantum computation with the surface code for the first time.

\section*{Acknowledgements}

We thank Austin Fowler for sampling the opinions of other researchers, for assistance locating references, for checking our ancilla opportunity cost estimates, and for comments that significantly improved this paper.
We thank Dave Bacon and Dmitri Maslov for comments that significantly improved this paper.

\bibliographystyle{plainnat}
\bibliography{citations}

\begin{thebibliography}{23}
\providecommand{\natexlab}[1]{#1}
\providecommand{\url}[1]{\texttt{#1}}
\expandafter\ifx\csname urlstyle\endcsname\relax
  \providecommand{\doi}[1]{doi: #1}\else
  \providecommand{\doi}{doi: \begingroup \urlstyle{rm}\Url}\fi

\bibitem[Amy et~al.(2013)Amy, Maslov, Mosca, and Roetteler]{Amy2013}
M.~Amy, D.~Maslov, M.~Mosca, and M.~Roetteler.
\newblock A meet-in-the-middle algorithm for fast synthesis of depth-optimal
  quantum circuits.
\newblock \emph{{IEEE} Transactions on Computer-Aided Design of Integrated
  Circuits and Systems}, 32\penalty0 (6):\penalty0 818--830, jun 2013.
\newblock \doi{10.1109/tcad.2013.2244643}.

\bibitem[Babbush et~al.(2018)Babbush, Gidney, Berry, Wiebe, McClean, Paler,
  Fowler, and Neven]{Babbush2018}
Ryan Babbush, Craig Gidney, Dominic~W Berry, Nathan Wiebe, Jarrod McClean,
  Alexandru Paler, Austin Fowler, and Hartmut Neven.
\newblock Encoding electronic spectra in quantum circuits with linear {T}
  complexity.
\newblock \emph{arXiv preprint arXiv:1805.03662}, 2018.
\newblock URL \url{https://arxiv.org/abs/1805.03662}.

\bibitem[Barenco et~al.(1995)Barenco, Bennett, Cleve, DiVincenzo, Margolus,
  Shor, Sleator, Smolin, and Weinfurter]{Barenco1995}
Adriano Barenco, Charles~H. Bennett, Richard Cleve, David~P. DiVincenzo, Norman
  Margolus, Peter Shor, Tycho Sleator, John~A. Smolin, and Harald Weinfurter.
\newblock Elementary gates for quantum computation.
\newblock \emph{Physical Review A}, 52\penalty0 (5):\penalty0 3457--3467, nov
  1995.
\newblock \doi{10.1103/physreva.52.3457}.

\bibitem[Barends et~al.(2014)Barends, Kelly, Megrant, Veitia, Sank, Jeffrey,
  White, Mutus, Fowler, Campbell, Chen, Chen, Chiaro, Dunsworth, Neill,
  O'Malley, Roushan, Vainsencher, Wenner, Korotkov, Cleland, and
  Martinis]{Bare13}
R.~Barends, J.~Kelly, A.~Megrant, A.~Veitia, D.~Sank, E.~Jeffrey, T.~C. White,
  J.~Mutus, A.~G. Fowler, B.~Campbell, Y.~Chen, Z.~Chen, B.~Chiaro,
  A.~Dunsworth, C.~Neill, P.~O'Malley, P.~Roushan, A.~Vainsencher, J.~Wenner,
  A.~N. Korotkov, A.~N. Cleland, and John~M. Martinis.
\newblock Superconducting quantum circuits at the surface code threshold for
  fault tolerance.
\newblock \emph{Nature}, 508:\penalty0 500--503, 2014.
\newblock \doi{10.1038/nature13171}.
\newblock arXiv:1402.4848.

\bibitem[Bravyi and Kitaev(1998)]{Brav98}
S.~B. Bravyi and A.~Yu. Kitaev.
\newblock Quantum codes on a lattice with boundary.
\newblock \emph{arXiv:quant-ph/9811052}, 1998.
\newblock URL \url{https://arxiv.org/abs/quant-ph/9811052}.

\bibitem[Brent and Kung(1982)]{brent1982}
Richard~P Brent and H-T\_ Kung.
\newblock A regular layout for parallel adders.
\newblock \emph{IEEE transactions on Computers}, \penalty0 (3):\penalty0
  260--264, 1982.
\newblock \doi{10.1109/TC.1982.1675982}.

\bibitem[Cuccaro et~al.(2004)Cuccaro, Draper, Kutin, and Moulton]{Cuccaro2004}
Steven~A. Cuccaro, Thomas~G. Draper, Samuel~A. Kutin, and David~Petrie Moulton.
\newblock A new quantum ripple-carry addition circuit, 2004.
\newblock URL \url{https://arxiv.org/abs/quant-ph/0410184}.

\bibitem[Dennis et~al.(2002)Dennis, Kitaev, Landahl, and Preskill]{Denn02}
E.~Dennis, A.~Kitaev, A.~Landahl, and J.~Preskill.
\newblock Topological quantum memory.
\newblock \emph{J. Math. Phys.}, 43:\penalty0 4452--4505, 2002.
\newblock \doi{10.1063/1.1499754}.
\newblock arXiv:quant-ph/0110143.

\bibitem[Draper et~al.(2004)Draper, Kutin, Rains, and Svore]{Draper2004}
Thomas~G. Draper, Samuel~A. Kutin, Eric~M. Rains, and Krysta~M. Svore.
\newblock A logarithmic-depth quantum carry-lookahead adder.
\newblock 2004.
\newblock URL \url{https://arxiv.org/abs/quant-ph/0406142}.

\bibitem[Fowler et~al.(2017)Fowler, Maslov, Jones, and
  Amy]{AustinDiscussionsAndEmails2017}
Austin Fowler, Dmitri Maslov, Cody Jones, and Matt Amy.
\newblock Private correspondence, Aug 2017.

\bibitem[Fowler et~al.(2012)Fowler, Mariantoni, Martinis, and
  Cleland]{Fowler2012}
Austin~G. Fowler, Matteo Mariantoni, John~M. Martinis, and Andrew~N. Cleland.
\newblock Surface codes: Towards practical large-scale quantum computation.
\newblock \emph{Physical Review A}, 86\penalty0 (3), sep 2012.
\newblock \doi{10.1103/physreva.86.032324}.

\bibitem[Gambetta et~al.(2017)Gambetta, Chow, and Steffen]{Gamb17}
J.~M. Gambetta, J.~M. Chow, and M.~Steffen.
\newblock Building logical qubits in a superconducting quantum computing
  system.
\newblock \emph{npj Quantum Information}, 3\penalty0 (2), 2017.
\newblock \doi{10.1038/s41534-016-0004-0}.
\newblock arXiv:1510.04375.

\bibitem[Horsman et~al.(2012)Horsman, Fowler, Devitt, and
  Van~Meter]{horsman2012}
Clare Horsman, Austin~G Fowler, Simon Devitt, and Rodney Van~Meter.
\newblock Surface code quantum computing by lattice surgery.
\newblock \emph{New Journal of Physics}, 14\penalty0 (12):\penalty0 123011,
  2012.
\newblock \doi{10.1088/1367-2630/14/12/123011}.

\bibitem[Howard and Campbell(2017)]{Howard2017}
Mark Howard and Earl Campbell.
\newblock Application of a resource theory for magic states to fault-tolerant
  quantum computing.
\newblock \emph{Physical review letters}, 118\penalty0 (9):\penalty0 090501,
  2017.
\newblock \doi{10.1103/PhysRevLett.118.090501}.

\bibitem[Jones(2013)]{Jones2013}
Cody Jones.
\newblock Low-overhead constructions for the fault-tolerant toffoli gate.
\newblock \emph{Physical Review A}, 87\penalty0 (2), feb 2013.
\newblock \doi{10.1103/physreva.87.022328}.

\bibitem[Kitaev et~al.(2002)Kitaev, Shen, and Vyalyi]{Kitaev2002}
Alexei~Yu Kitaev, Alexander Shen, and Mikhail~N Vyalyi.
\newblock \emph{Classical and quantum computation}.
\newblock Number~47. American Mathematical Soc., 2002.
\newblock \doi{10.1090/gsm/047}.

\bibitem[Lahtinen and Pachos(2017)]{Laht17}
V.~Lahtinen and J.~K. Pachos.
\newblock A short introduction to topological quantum computation.
\newblock \emph{arXiv:1705.04103}, 2017.
\newblock URL \url{https://arxiv.org/abs/1705.04103}.

\bibitem[Lekitsch et~al.(2017)Lekitsch, Weidt, Fowler, M{\o}lmer, Devitt,
  Wunderlich, and Hensinger]{Leik17}
B.~Lekitsch, S.~Weidt, A.~G. Fowler, K.~M{\o}lmer, S.~J. Devitt, C.~Wunderlich,
  and W.~K. Hensinger.
\newblock Blueprint for a microwave trapped-ion quantum computer.
\newblock \emph{Science Advances}, 3\penalty0 (2):\penalty0 e1601540, 2017.
\newblock \doi{10.1126/sciadv.1601540}.
\newblock arXiv:1508.00420.

\bibitem[Muñoz-Coreas and Thapliyal(2017)]{Coreas2017}
Edgard Muñoz-Coreas and Himanshu Thapliyal.
\newblock T-count optimized design of quantum integer multiplication, 2017.
\newblock URL \url{https://arxiv.org/abs/1706.05113}.

\bibitem[Nielsen and Chuang(2009)]{Nielsen2009}
Michael~A. Nielsen and Isaac~L. Chuang.
\newblock \emph{Quantum Computation and Quantum Information}.
\newblock Cambridge University Press, 2009.
\newblock \doi{10.1017/cbo9780511976667}.

\bibitem[Raussendorf and Harrington(2007)]{Raus07}
R.~Raussendorf and J.~Harrington.
\newblock Fault-tolerant quantum computation with high threshold in two
  dimensions.
\newblock \emph{Phys. Rev. Lett.}, 98:\penalty0 190504, 2007.
\newblock \doi{10.1103/PhysRevLett.98.190504}.
\newblock arXiv:quant-ph/0610082.

\bibitem[Raussendorf et~al.(2007)Raussendorf, Harrington, and Goyal]{Raus07d}
Robert Raussendorf, Jim Harrington, and Kovid Goyal.
\newblock Topological fault-tolerance in cluster state quantum computation.
\newblock \emph{New Journal of Physics}, 9\penalty0 (6):\penalty0 199, 2007.
\newblock \doi{10.1088/1367-2630/9/6/199}.

\bibitem[Schlosser et~al.(2011)Schlosser, Tichelmann, Kruse, and Birkl]{Schl11}
Malte Schlosser, Sascha Tichelmann, Jens Kruse, and Gerhard Birkl.
\newblock Scalable architecture for quantum information processing with atoms
  in optical micro-structures.
\newblock \emph{Quantum Information Processing}, 10\penalty0 (6):\penalty0 907,
  2011.
\newblock \doi{10.1007/s11128-011-0297-z}.
\newblock 1108.5136.

\end{thebibliography}

\end{document}